# High-resolution boreal winter precipitation projections over Tropical America from CMIP5 models

Reiner Palomino-Lemus[1,2], Samir Córdoba-Machado[1,2], Sonia Raquel Gámiz-Fortis[1], Yolanda Castro-Díez[1], María Jesús Esteban-Parra[1]

[1] Department of Applied Physics, University of Granada, Granada, Spain

[2] Technological University of Chocó, Colombia

Sonia Raquel Gámiz-Fortis (ORCID ID: 0000-0002-6192-056X)
Yolanda Castro-Díez (ORCID ID: 0000-0002-2134-9119)
María Jesús Esteban-Parra (ORCID ID: 0000-0003-1350-6150)

(*) Corresponding author address:

María Jesús Esteban Parra

Departamento de Física Aplicada

Facultad de Ciencias

Universidad de Granada

Campus Fuentenueva s/n

18071-Granada. Spain.

E-mail: esteban@ugr.es

Phone: +34 958 240021

Fax: +34 958 243214

**ABSTRACT**

Climate-change projections for boreal winter precipitation in Tropical America has been addressed by statistical downscaling (SD) using the principal component regression with sea-level pressure (SLP) as the predictor variable. The SD model developed from the reanalysis of SLP and gridded precipitation GPCC data, has been applied to SLP outputs from 20 CGMS of CMIP5, both from the present climate (1971-2000) and for the future (2071-2100) under the RCP2.6, RCP4.5, and RCP8.5 scenarios. The SD model shows a suitable performance over large regions, presenting a strong bias only in small areas characterized by very dry climate conditions or poor data coverage. The difference in percentage between the projected SD precipitation and the simulated SD precipitation for present climate, ranges from moderate to intense changes in rainfall (positive or negative, depending on the region and the SD GCM model considered), as the radiative forcing increases from the RCP2.6 to RCP8.5. The disparity in the GCMs outputs seems to be the major source of uncertainty in the projected changes, while the scenario considered appears less decisive. Mexico and eastern Brazil are the areas showing the most coherent decreases between SD GCMs, while northwestern and southeastern South America show consistently significant increases. This coherence is corroborated by the results of the ensemble mean which projects positive changes from 10ºN towards the south, with exceptions such as eastern Brazil, northern Chile and some smaller areas, such as the center of Colombia, while projected negative changes are the majority found in the northernmost part.

**Keywords**: boreal winter precipitation; climate projections; Tropical America; statistical downscaling; CMIP5 GCMs.

# 1. INTRODUCTION

Producing reliable estimates of changes in precipitation at local and regional level remains a major challenge in climate science, as it is a key aspect for planning adaptation and mitigation measures in order to reduce the negative impacts of the climate change in vulnerable regions (Giorgi et al. 2001; Christensen et al. 2007). The tropical American region, because of its meteorological and climatological characteristics, has received a special attention from the scientific community over recent decades. Unique environments, such as the Amazonia (the largest tropical rainforest on the planet), the Andes Mountains (with steep slopes), the desert of Atacama in Chile, the arid region of northeastern Brazil, the extreme west of Peru and Ecuador, the biodiversity of western Colombia and western Central America, the migration of the Intertropical Convergence Zone (ITCZ), the South American Monsoon System, among others, that interact in a complex superposition of physical processes at diverse spatio-temporal scales, determine the meteorological and climatological aspects of Tropical America, constituting a fundamental component of the global system. In turn, the main features of atmospheric circulation are associated with precipitation in the region, which directly and indirectly affect the economy, ecosystems, and society (Alexander et al. 2002; Barsugli and Sardeshmukh 2002). The Fifth Assessment Report of the Intergovernmental Panel on Climate Change (IPCC AR5 2013a, 2013b) suggests both increases and decreases in rainfall for Central and South America by 2100, depending on the region, although with high uncertainties due to high discrepancies between different General Circulation Models (GCMs) projections. According to Magrin et al. (2014), changes in agricultural production, with consequences for food supply, associated with climate change, are expected to show significant spatial variability in Central and South America (Marengo et al. 2010). The increase in agricultural production and intensive land use could lead to desertification, water pollution, erosion, and negative effects on biodiversity and health. For this reason, the study of climate change in this area constitutes a vital objective for the socio-economic development of the region.

Dynamic (DD) and statistical (SD) downscaling methods (Schmidli et al., 2006; Zorita and von Storch 1999; von Storch et al. 2000) are often used to reduce the gap between the coarse resolution of GCMs and the information at higher spatial resolution (Grotch and MacCraken 1991; von Storch et al. 1993; Wilby and Wigley 1997; Xu 1999). While the DD methods use a high-resolution regional climate model nested in a GCM, the SD is performed by looking for empirical statistical relationships between large scale atmospheric predictors and regional scale variables (Wood et al. 2004; Yang and Wang 2012), assuming that these will be maintained over time under future climate conditions. The SD presents the added benefit of low computational cost versus DD methods. There are uncertainties in the projections associated with both methodologies, such as the parameterizations (in the DD) or the predictors choice (in the SD) (Frost et al., 2011; Bae et al., 2011; Wilby and Wigley 2000). Little consensus exists on which predictors are more appropriate, although variables related to atmospheric circulation, such as level pressure (SLP) are widely used, due to their availability from both observational and GCM output data. One of the most frequently used approaches for developing SD models is the principal component regression (PCR), which is based on the principal component analysis (PCA) to reduce the dimensionality of the predictor data (Preisendorfer 1988; Jolliffe 2002; Wilks 2006). According to the use of principal components (PCs) as predictors, the SD model generated by PCR, which takes into account the interactions between predictands and observed predictors, is applied to results from the GCM outputs representing climate change projections (Wilks 2006; Li and Smith 2009; Eden and Widmann 2014). However, before the SD model can be applied to project changes in rainfall for the end of the century, an evaluation of the ability of the SD model to reproduce the present climate should be performed. In any case, the climate change estimations at the regional scale are affected by different uncertainties coming from the different GCMs, scenarios, and the downscaling method itself selected.

The use of several GCMs and scenarios is important to reduce some of these uncertainties (Wilby and Harris 2006; Maurer 2007). Thus, one way to analyze the uncertainty is to work with a multimodel ensemble (Palmer et al. 2005), which provides a probability distribution of possible future values (Harris et al. 2010). Some studies have demonstrated that simulation errors and uncertainties using individual GCMs could be reduced by the use of the ensemble mean of the members for multi-model projections. This is true for studies concerning the verification of seasonal forecasts (Palmer et al. 2004; Hagedorn et al. 2005), present-day climate from long-term simulations (Lambert and Boer 2001) or climate change projections (Nohara et al. 2006). So, the ensemble average usually reproduces the observations better than do individual models (Wallach et al. 2016).

In the current literature few works attempt projections of climate change in Tropical America, most research being more focused on particular regions such as Brazil, Colombia or southern South America (Ramírez et al. 2006; Solman and Nuñez 1999; Mendes and Marengo 2010; Teichmann et al. 2013, Palomino-Lemus et al. 2015). Thus, there is a clear need for the study of climate change in Tropical America.

The present work takes into account all the previous considerations and has a primary aim to obtain climate change projections for the boreal winter precipitation of Tropical America, during the period 2071-2100. For this, the precipitation has been statistically downscaled, using as predictor the SLP from the tropical Pacific through the PCR technique. Once the skill of the SD model developed was demonstrated for simulating the rainfall of the region under the present climate, this was applied to the SLP simulations of 20 GCMs selected from the Coupled Model Intercomparison Project Phase 5 (CMIP5, Taylor et al. 2012), for three representative concentration pathways, RCP2.6, RCP4.5, and RCP8.5. The study is structured as follows. Section 2 describes the datasets used, Section 3 explains the methodology, Section 4 displays the results, and Section 5 presents the concluding remarks.

## 2. DATA

For this study, the observational precipitation dataset from the Global Precipitation Climatology Centre, GPCC version 6.0 (Schneider et al. 2014) was used. The boreal winter precipitation, composed by the averaged December, January, and February (DJF) rainfall over the 61-yr period, from 1950 to 2010, was generated from GPCC data. The time series of winter rainfall corresponding to the grid points of the study region [30°N–30°S, 120°W–30°W] (Figure 1), with a spatial resolution of 0.5°×0.5°, were used as the predictand in the process of building a SD model, using principal component regression (PCR) method, to simulate the boreal winter precipitation for the period 1950-2010.

As a predictor variable, the mean monthly sea level pressure (SLP) data available from the National Center for Environmental Prediction-National Center for Atmospheric Research (NCEP-NCAR reanalysis project), which has a spatial grid resolution of 2.5°×2.5° (Kalnay et al. 1996), was used, covering a more extensive area [30°S–30°N, 180°W–30°W] for the same period 1950-2010.

In addition, SLP outputs from 20 GCMs, taken from the CMIP5 (Taylor et al. 2012), were used. These models were chosen for their accurate reproduction of the SLP variability modes (Palomino-Lemus et al. 2015). The model data include simulations with historical atmospheric concentrations and future projections for the representative concentration pathways RCP2.6, RCP4.5, and RCP8.5 (Moss et al. 2010; Taylor et al. 2012). The historical experiments cover the period from 1850 to 2005. In the present study, the period 1971-2000 was used as representative of present climate, while, for the future climate, the period 2071-2100 was considered. Table 1 shows these 20 GCMs, labeled from (a) to (t) for their identification, and their principal features. In all the cases, the *run1* of the simulations for historical climate was used.

## 3. METHODOLOGY

Statistical downscaling is a process consisting of a double step. First, a search was made of relationships between the local climate variables and the large-scale predictors (winter precipitation and SLP, respectively, in our case). Second, the relationships found were applied to the GCMs outputs to develop a SD model.

A key point to take into account in this process is the multicollinearity between data subset, which could be a serious problem when a statistical regression model has a great number of input data, because the number of estimated regression coefficients can be very large, resulting in misleading estimates of the regression equation (Draper and Smith 1981; Jolliffe 2002). To address the problems associated with multicollinearity, we used biased regression estimators, such as the principal components regression (PCR) method, as frequently suggested. A detailed description of this methodology can be seen in Palomino-Lemus et al. (2015).

In this work the spatio-temporal variability of SLP reanalysis data from NCEP was analyzed by PCA using the covariance matrix (Preisendorfer 1988). Empirical orthogonal functions (EOFs) and principal components (PCs) that account for a high percentage of explained SLP variance, presenting significant correlations with the winter precipitation in the study area, were selected. For an assessment of the robust correlations between the main leading SLP PCs and DJF precipitation, the non-parametric bootstrap technique (Stine 1985; Li and Smith 2009) was used, identifying significant correlations at the 95% confidence level. When the main PCs of SLP were selected, the PCR method was applied to model the winter precipitation following the scheme proposed by Li and Smith (2009). The periods 1950-1993 and 1994-2010 were used for calibration and validation, respectively. The Bootstrap with replacement was applied to provide estimates of the statistical errors. Afterwards, the statistical model built for each grid point was recalibrated using the total observational period (1950-2010), allowing us to consider the most recent variability of the fields in the regression model, and finally, to generate the definitive SD model.

The skill of the different GCMs to simulate the DJF rainfall in the Tropical America for present climate (1971-2000) was studied by computing the differences between the simulated and observed precipitation values. Lastly, to project DJF precipitation in the area for the period 2071-2100, the SD model, was applied to the SLP outputs from 20 GCMs under the RCP2.6, RCP4.5, and RCP8.5 scenarios. The non-parametric rank sum test of Wilcoxon-Mann-Whitney (von Storch and Zwiers 2013) was applied to analyze the significance of the changes projected.

Finally, to take the advantage of reducing simulation errors and uncertainties (Lambert and Boer 2001; Palmer et al. 2004; Hagedorn et al. 2005; Nohara et al. 2006), we calculated the projected precipitation changes under the three scenarios using the arithmetic ensemble mean of the 20 SD GCM outputs.

## 4. RESULTS

### 4.1 Spatio-temporal SLP modes and their relationship with precipitation

A PCA applied to the DJF SLP reanalysis data in the period 1950-2010 identifies 10 leading modes of variability that explain 88.8% of the total variance. Figure 2 shows the spatial patterns (EOFs) of these modes and their corresponding PC series.

The first mode of variability (EOF1) explains 31.5% of the total variance of the SLP data, and is characterized by the presence of a dominant pattern of positive correlations that represents the variability of almost the entire region of tropical Pacific Ocean included in this study, with a strong positive correlation center located around the 150ºW-10ºS, stretching to the northern tropical Atlantic. The second mode (EOF2), which explains 16.9% of the SLP variance, exhibits two well-defined action centers, one with positive correlations located in the northwestern edge of the study area, and the other with negative correlations extending from the Gulf of Mexico, covering all Central America to approximately 150ºW. EOF3 (12.3% of variance), shows a spatial pattern with a strong core of positive correlations in the northeast, centered around 15°N-40ºW, which spreads, though weakened, throughout northern South America, to northern Chile. Additionally a gradient of negative correlations, which is distributed from the south end to the 10ºS, between 170ºW and 90ºW, also appears. EOF4 (8.8% of variance) shows two negative centers located in the west Pacific and South America, respectively, along with a weaker positive center covering the Gulf of Mexico, the Florida peninsula and most of the Caribbean islands. EOF5 (8.8% of variance) to EOF10 jointly account for 19.3% of the SLP variance and show different action centers over the study region with weaker factor loadings.

To explore the physical meaning of these variability modes, we analyzed the correlations between their corresponding PC series (also shown in Figure 2) and several teleconnection indices. The results show that the first PC series is related to the ENSO and SOI indices, the highest correlation coefficient being for bivariate ENSO index (BEST, Smith and Sardeshmukh 2000) (r = -0.71), followed by El Niño4 (r = -0.68) and El Niño3.4 (r = -0.65) indices, all significant at 95% confidence level. PC2 is strongly correlated with the Western Pacific (WP) index (r = 0.80), and with El Niño1+2 index (r = 0.53). PC3 is related to the Atlantic SST, showing the highest negative correlations with the Atlantic Meridional Mode (AMM, Chiang and Vimont 2004) (r = -0.63), followed by the Atlantic Tripole SST EOF (ATLTRI, Deser and Timlin 1997) (r = -0.54) and the Tropical Northern Atlantic (TNA, Enfield et al. 1999) (r = -0.52) indices. The PC4 shows significant correlation with the Pacific SST, being the highest coefficient with the Western Hemisphere Warm Pool (WHWP, Wang and Enfield 2001) (r = -0.53) index.

For the analysis of the relationships between the SLP and precipitation, Figure 3 shows the spatial distribution of the correlation coefficients between DJF precipitation data and each time PC series associated with the 10 main modes of variability of DJF SLP. Only statistically significant results at 95% confidence are colored. Additionally, the percentage of area covered by these significant correlations is also shown. The correlation map for the PC1 (Figure 3a) clearly presents significant correlations in an extended area of the region, with significant correlations covering about 40.9% of the region, being the SLP PC which correlates most extensively with the precipitation of the study region. The correlation map for this PC1 is dominated by a broad band of positive correlations that starts from the southwest and northern Brazil and extends to northern Nicaragua. In this area, two main centers have the highest values of positive correlation (above 0.6), located northwest of the Andes in Colombia, and the other in northern Brazil, reaching the east of Venezuela, and entirely covering Guiana, Surinam, and French Guiana. These positive correlations show the influence of the first DJF SLP mode of variability on DJF precipitation in these regions. In addition, significant negative correlations also appear, with values of up to -0.5, especially in Mexico, and slightly weaker in southeastern Brazil, in Paraguay, and in northeastern Argentina. Since PC1 is related mainly to the ENSO phenomenon, this result indicates a clear association between ENSO and DJF precipitation variability in the area of Tropical America.

The next DJF SLP mode of variability that presents the second highest percentage (31.1%) of continental area with significant correlations with precipitation, is associated with the SLP PC3. The spatial correlation map (Figure 3c) shows a pattern similar to that of the PC1 (Figure 3a), with certain differences, but with opposite sign correlations. It has negative correlations in northern South America, stretching from Colombia to French Guiana, while positive correlations are located in northern Mexico, the Yucatan Peninsula and central Brazil. PC4 follows the third mode in percentage of area with significant correlations (Figure 3d), with 24.6%, and is characterized by the presence of lower and more localized correlation values. Regionally, it presents significant positive correlations with precipitation in Venezuela, Guiana, Surinam, and French Guiana, and negative in northeastern Argentina and southern end of Brazil.

In addition, the correlation between DJF SLP PC2 and DJF precipitation (Figure 3b), presents, generally low values, showing significant positive correlations only in the Florida peninsula, some Caribbean islands and western Ecuador; and negative ones in Guiana, Surinam and at the mouth of the Amazon River in northern Brazil. These areas represent only 16% of total area.

Moreover, the rest of DJF SLP PCs (PC5, PC8, PC7, PC10, PC9, and PC6) have lower percentages of areas with significant correlations (14.7%, 14.7%, 12.4%, 11.5%, 10.8%, and 8.9%, respectively). Note the PC5 correlations (Figure 3e), for which there are two centers of significant correlations with opposite signs located to the east of Brazil, and in southern Brazil, and in southern Paraguay, as well as PC8 (Figure 3h), for which a large center to the east of Brazil with significant negative correlations is shown. The rest of PCs show weaker correlations with precipitation, identifying localized regions scattered over the area of study.

## 4.2 Statistical downscaling model

After the analysis of the relationships between SLP and precipitation, the aim was to develop a robust statistical model that would provide the downscaled precipitation for each grid point from the large-scale SLP field. The PCR method was used to build the statistical downscaling (SD) model for DJF rainfall, using the PC series corresponding to the first 10 modes of variability of DJF SLP NCEP reanalysis data as predictor variables, and the observed gridded DJF precipitation as predictands. As mentioned above, the training period 1950-1993 was used as calibration period, and the period 1994-2010 to validate the model.

Figure 4 shows the spatial distribution of the correlation coefficients between observed DJF precipitation data and the generated with the SD model for each grid point during the calibration (1950-1993) and validation (1994-2010) periods (Figure 4a and 4b, respectively). The highest correlations ($r > 0.8$) for the validation period are found in southern Central America, in the northwestern regions of Colombia and Ecuador, and in the northwestern end of Peru. There are also high correlations extending from eastern Venezuela to northern Brazil, covering Guiana, Surinam, and French Guiana. Additionally, strong correlation values appear in many scattered areas, such as Florida and south of the study area. On the other hand, comparing the calibration period with the validation one, lower correlation coefficients are found for the latter area, mainly from southern Mexico (through the Yucatan Peninsula) to Honduras. Lower values are also appreciated southeast of Colombia, northern Venezuela and a vast area over the center of South America.

The relative root mean square error (RMSE) was used to quantify the differences between observed and simulated precipitation as well as to assess the stability of the SD model. The spatial distribution of the percentage of RMSE during the calibration and validation periods is shown in Figure 5a and 5b, respectively, reflecting great similarity between the two periods. Some regions have relatively large errors, such as Chile, coastal Peru, southwestern Bolivia, and Mexico, all registering low precipitation values. Generally, errors are lower on the southern half of the study area, while in the north the opposite happens.

For a direct comparison between simulated and observed precipitation values at each grid point, Figure 6 depicts the spatial distribution of the observed (Figure 6a) and simulated DJF precipitation (Figure 6b) for the validation period (1994-2010), as well as the spatial distribution of the percentage differences between the two fields (Figure 6c). This comparison shows that the SD model provides a good representation of the average DJF rainfall field, with very small differences between observed and simulated values. Moreover, the maximum values of rainfall in the region, over relatively small areas in western Colombia, southeastern Peru, and central Bolivia, are properly reproduced. The major discrepancies are associated with very dry areas or without information, such as the western edge of South America or the Pacific coast of Mexico, where both underestimations and overestimations of precipitation are appreciated.

### 4.3 Simulated DJF precipitation for present climate

After assessing the ability of the SD model, we recalibrated it using the complete period 1950-2010. Figure 7 presents the spatial distribution of the correlation coefficients between observed DJF precipitation data and the SD modeled values during the period of recalibration (Figure 7a), as well as the ones estimated from the SD model for the period 1971-2000 (Figure 7b), which will be used as reference period to characterize precipitation in the present climate. For both the calibration (1950-1993, Figure 4a) and recalibration (1950-2010, Figure 7a) periods, the SD model shows the same spatial correlation pattern. For the period 1971-2000, correlations for certain relatively large areas prove poorer, while in more limited and scattered areas the correlation improves, but remaining essentially the same spatial configuration of the correlation as for the other periods. Figure 7c shows the percentage differences between the observed DJF precipitation and the results from SD modeled one using the SLP, for the period 1971-2000. Only a small very dry area over the northwest of Chile presents remarkable bias.

After recalibrating the SD model for the complete period 1950-2010, and assess its ability to reproduce the precipitation in each grid point, this was applied to SLP data derived from 20 GCMs, selected from CMIP5 (Table 1) for both present climate (1971-2000) and future climate (2071-2100) under the RCP2.6, RCP4.5, and RCP8.5 scenarios.

Figure 8 shows the percentage of the differences between the SD precipitation from 20 GCMs and the observed DJF precipitation for 1971-2000 period. Additionally, the statistical significance at 95% confidence level of these differences was estimated using the Wilcoxon-Mann-Whitney bilateral rank sum test. The results show that, generally, there are no statistically significant differences for a large number of models, indicating that the SD model applied to the SLP outputs of these GCMs has a high ability to faithfully reproduce the precipitation field. However, the simulations performed directly by using non-downscaled outputs of GCMs (Figure 9) strongly distort the precipitation field, since they are able to reproduce neither the values nor the spatial distribution of precipitation. Note that the area with significant differences (Figure 8) is on average (considering the SD of all models) only 16.79% for the period 1971-2000. Therefore, the

SD applied to the 20 GCMs accurately reproduces the highest and lowest values of the rainfall in most of the study area. Furthermore, these SD precipitation values (not shown) are very close to those observed, showing spatial patterns very similar to the observed ones.

The results of Figure 8 also reveal that, although the SD model successfully reproduces the most important spatial patterns of DJF precipitation in the study area, significant deficiencies are evident for simulations made with outputs from MIROC-ESM (p) and GISS-E2-R (k), followed by GFDL-CM3 (j), with a percentage of the area showing significant differences higher than 20%. In particular, for GISS-E2-R model (Figure 8k), SD overestimates by more than 60% the observed rainfall in areas located above 20°N, covering Mexico. Meanwhile, for the MIROC-ESM (Figure 8p), differences in percentage strongly underestimate precipitation in Mexico ($< -90\%$).

### 4.4 Projected changes in DJF precipitation

Figures 10, 11, and 12 show the percentage of changes in projected (2071-2100) DJF rainfall compared to the present (1971-2000) SD precipitation for each GCM under the RCP2.6, RCP4.5, and RCP8.5 scenarios, respectively. The statistical significance of the projected precipitation changes, as previously, has been estimated by using the bilateral rank sum test of Wilcoxon-Mann-Whitney. As can be seen, for the 20 projected predictions in general, the RCP4.5 and RCP8.5 scenarios show large areas with significant changes. For the RCP2.6 scenario (Figure 10), projected results reflect a predominance of very moderate decreases in rainfall, these being significant in some models. The extent of the area affected by significant changes varies from 2.56% for the SD CSIRO-Mk3.6 (Fig. 10g) to 57.91% for SD HadGEM2-ES (Fig. 10m). The area with most consistent changes between the SD GCMs is eastern Brazil (around 10°S, 40°W), particularly intense (declines of more than 80%) in SD CanESM2 (Fig. 10c) and SD GFDL-CM3 (Fig. 10j) models. Some models also show a sharp decline in the Chilean Andes. Northern Mexico also presents significant declines from some SD models (around 30% or higher in some areas), while the southwestern Mexican coastal area shows increases (over 60%) for several SD GCMs.

As radiative forcing increases, the extent of the area with significant changes in precipitation also increases (Fig. 11 and 12). For example, for RCP8.5 (Fig. 12) the minimum extension with significant changes exceeds 40% (SD MPI-ESM-LR model, Fig. 12q, and SD MPI-ESM-MR model, Fig. 12r), reaching 80% is some case (SD NorESM1-ME model, Fig. 12t). This latter SD model also presents a greater surface area with significant changes under the RCP4.5 scenario (Fig. 11t). For this RCP4.5 scenario (Fig. 11), some models have fewer areas with significant changes than for the RCP2.6 one (SD IPSL-CM5A-MR, Fig. 11n; SD MPI-ESM-MR, Fig. 11r; and especially the SD BCC-ESM1.1, Fig. 11b). In addition, there are more changes towards a decline in rainfall, which become very marked again in eastern Brazil (SD CanESM2, Fig. 11c, and SD GFDL-CM3, Fig. 11j), and Mexico (SD MIROC5, Fig. 11o, and SD NorESM1-ME, Fig. 11t). However, the changes shown are less consistent in some areas, such as northern South America, where some models show increases (SD CNRM-CM5, Fig. 11f, and SD

GISS-E2-R, Fig 11k) and other reductions (SD FGOALS-g2, Fig 11h, and SD HadGEM2-AO, Fig. 11l), or even opposing trends in relatively nearby areas ( SD MRI-CGCM3, Fig. 11s).

For RCP8.5 (Fig. 12), the SD of 13 GCMs show strongly significant declines (above 30%) in most of Mexico, especially in the north, reaching over -90% in some cases (SD MIROC5, Fig. 12o, and SD NorESM1-ME, Fig. 12t). Eastward of Brazil (10ºS, 40ºW), similar results appear for 13 GCMs, showing significant decreases. In the northwest of South America (west of Colombia) simulations (for 12 GCMs), showing significant increases in precipitation predominate, in the northernmost part reaching an 80% increase (SD HadGEM2-ES, Fig. 12m).

To identify how robust the projected precipitation changes are, we have studied the coherence between the results of the 20 SD GCMs by calculating the percentage of them that agree in the sign of projected precipitation change at each grid point of the study area. Only coherence values higher than 55% are shown. The Figure 13 depicts these results, showing that the projected precipitation changes have great coherence between the 20 SD models in most of the area, with positive or negative changes depending on the region and the scenario considered. The areas that are consistently affected by increased or decreased rainfall are spread as the radiative forcing increases, except for the region between Venezuela and Guiana, where there is a light loss of coherence. In general, there are wide spatial areas with coherence higher that 80%. Note for example the border region between Colombia, Ecuador, and Peru, the border between Brazil and Paraguay and the southern tip of Brazil, with coherent positive projected changes. Meanwhile, the diagonal band between the northwestern Brazil to the east coast of Brazil located around 20°S-40°W, the border between Bolivia, Chile, and Argentina, and an extended area covering Mexico and Central America, present coherent negative projected changes. The high coherence (higher than 90% in some grid points) is remarkable between the SD GCMs in the narrow area of Central America, where almost all the models are able to discriminate between positive changes in the Pacific coast and negative ones in the Atlantic coast.

The coherence found between the sign of the projected precipitation changes for 20 SD GCMs provides the base to generate multimodel ensemble projections. The projected precipitation changes under the three scenarios considered were calculated from the arithmetic ensemble mean of the 20 SD GCM outputs. Figure 14 shows the percentage of changes in projected (2071-2100) DJF rainfall compared to the present (1971-2000) SD precipitation for the ensemble multi-model mean under the RCP2.6, RCP4.5 and RCP8.5 scenarios, respectively. The statistical significance of the projected precipitation changes, as before, was estimated by the Wilcoxon-Mann-Whitney test. The results show that the projected changes were significant in most of the study area, covering from 66.27% under the RCP2.6 scenario, up to 83.95% under the RCP8.5. Projected changes are mostly moderate, covering extended regions with coherent sign, even under the scenario of highest radiative forcing. For all scenarios, areas with increased precipitation predominate over those where a decline is projected, although the prevalence increases with the radiative forcing considered, becoming 48.38% vs. 35.57% under the RCP8.5 scenario. Note the sharp increase projected in some parts of the Pacific coast, especially in southern

Mexico, Peru, and Chile, as well as the sharp decline in parts of Colombia, Venezuela, on the border between Brazil and Guiana, and areas of Chile.

## 5. CONCLUDING REMARKS AND DISCUSSION

The main goal of this work was to get climate change projections for boreal winter precipitation in Tropical America. For this, we developed a precipitation SD model for each grid point of the area by PCR technique using as predictors the SLP PCs series of NCEP data, and the observed gridded DJF precipitation as predictands. These predictors were rigorously selected according to the significance of their correlations with the observed precipitation field. Climate variability modes related to ENSO phenomenon can satisfactorily describe the precipitation in many areas of South America (Barros et al. 2000; Grimm et al. 2002; Tedeschi et al. 2013; Córdoba-Machado et al. 2015a, 2015b). For example, for Colombia precipitation these latter authors showed that the variability in the tropical Pacific SST, including El Niño and El Niño Modoki, is sufficient to reproduce and predict seasonal rainfall. El Niño phenomenon leads the variability of precipitation in much of the study region through its influence on the circulation of Walker, whose variations are reflected in the SLP field, this mode being particularly associated with the PC1 taken from the PCA applied to the tropical Pacific SLP. In addition, other patterns associated with the variability of the SLP on the tropical American continent and over the tropical Atlantic can also help in describing the behavior of precipitation in various areas of the tropical America, such as the Panama High or the northeastern Brazil Low pressure system. Moreover, some of the SLP PCs series analyzed in this study reflect the influence of certain extra-tropical Atlantic patterns, such as the Atlantic Meridional Mode, the Tripolar Atlantic SST or the Tropical Northern Atlantic pattern, whose contribution to the SD model could also be significant. So, in accordance with our results, other papers have shown that during the boreal winter (DJF), most of the moisture arriving to Central and South America comes from the Atlantic (Hoyos et al, 2017). In this sense, the ability of the SD model to predict the precipitation comes from the inclusion of these climate variability modes through their corresponding PCs.

In general, the SD model shows proper performance over large areas with small domains with major bias, particularly for the validation period (1994-2010). This may be due to the unreliable coverage of the GPCC data in certain areas (e.g. forest areas of the Amazon and Orinoco and Andes) in recent years, or regions characterized by very dry climate conditions (e.g. western edge of South America). These results are consistent with those reported by Eden et al. (2012) and Eden and Widmann (2014), who found bias greater than 10% in most of the tropics and in areas where the quality of the observation network is poor. However, SD model can properly reproduce the maximum values of rainfall in the region in western Colombia, southeastern Peru, or central Bolivia.

For present climate, while the simulations performed directly using GCM outputs are unable to reproduce the distribution of the precipitation field, there are no statistically significant differences between the observed DJF precipitation and the simulated one using the SD model for many GCMs. We find that, on average, the areas with significant differences represent only 16.79% of the complete region. Thus, the SD model applied to the selected GCMs can accurately reproduce the DJF precipitation field throughout most of the study area.

The high-resolution climate simulations projected for the end of this century have been evaluated using the difference in percentage between the projected SD precipitation for the period 2071-2100 and the simulated SD precipitation for the period 1971-2000. Results show positive or negative differences depending on the region and the SD GCM model considered. In general, these changes in rainfall range from very moderate to intense as the radiative forcing increases from the RCP2.6 to RCP8.5. Major sources of uncertainty in the projected precipitation changes for the end of the century seem to come from the disparity in the GCMs outputs, being less sensitive to the scenario considered. The results of the coherence between models shows that three northwest-to-southeast bands can be differentiated throughout the region, alternating projected changes in increased and decreased precipitation. Central and southeastern Brazil, Mexico and Guatemala are the areas showing the most consistent decrease changes between SD GCMs, while for the northwest and southeast of South America simulations showing significant increases predominate.

The mean ensemble shows regions having projected significant increases and significant decreases. While the percentage of area presenting negative significant changes is very similar for the three RCPs (from 32.06% to 35.74%), the percentage relative to significant positive changes is higher as the radiative forcing intensifies (ranging from 34.21% for the RCP2.6 to 48.38% for the RCP8.5). Basically, positive projected changes are found from 10ºN latitude to the south, with exceptions such as eastern Brazil, northern Chile and smaller areas such as the center of Colombia, while negative projected changes appear mostly in the northernmost part. The coherence of our results essentially agrees with the findings of Sánchez et al. (2015). Most of the simulations in this paper and in the present work show a precipitation decrease in the east and some interior parts of Brazil, as well as increases in the coast of Ecuador and Bolivia in addition to northern Argentina, Paraguay and southern Brazil, although Sánchez et al. (2015) used different GCMs, dynamical downscaling, and the A1B scenario. Chou et al. (2014), in their study of assessing the climate change over South America using dynamical downscaling, projected a reduction of DJF precipitation in a large area that extends from northwestern to southeastern South America, also especially important towards the end of the century and for the RCP8.5 in southeastern Brazil. However, comparing the results found in the present work with those reported by other authors is problematic because of the differences between regions, periods, seasons, GCMs, and scenarios analyzed.

Few studies have used the statistical downscaling over Tropical America, being more focused on the climate of some regions of Brazil or in the southern part of South America (Johnson et al. 2014; Valverde Ramírez et al. 2006; Solman and Nuñez 1999; Mendes and Marengo 2010). Hence the present study is novel for being one of the few papers devoted to obtain future rainfall projections at the regional scale for the Tropical America using CMIP5 models. Additionally, the statistical downscaling method developed in this work accurately reproduces the precipitation at the local scale for the study region, being, therefore, a useful technique for climate change studies, with the advantage of minimal computation requirement. Therefore the results of this work could be useful for the climate change mitigation purposes in this area.

**ACKNOWLEDGEMENTS**

Technological University of Chocó (UTCH) and COLCIENCIAS-Colombia by supported to R. Palomino-Lemus and S. Córdoba-Machado under a scholarship. The Spanish Ministry of Economy and Competitiveness, with additional support from the European Community Funds (FEDER), project CGL2013-48539-R and the Regional Government of Andalusia, project P11-RNM-7941, which had financed this study. We thank anonymous reviewers for valuable comments on the manuscript.

**Figure captions**

Figure 1: a) Region used for the precipitation study. b) Topographical features of the region of interest.

Figure 2. Loading factors for the 10 leading variability modes of the DJF SLP reanalysis data for the period 1950–2010 and their corresponding PC series.

Figure 3. Spatial correlation patterns between gridded DJF precipitation and the 10 leading PCs from NCAR DJF SLP. Only statistically significant results at 95% confidence are colored, and the percentage of area covered by these patterns is also shown.

Figure 4. Spatial distribution of the correlation coefficients between observed DJF precipitation values and simulated one by the SD model for each grid point during: a) calibration (1950-1993), and b) validation (1994-2010) periods.

Figure 5. Spatial distribution of the percentage of RMSE between observed DJF precipitation values and simulated one by the SD model for each grid point during: a) calibration (1950-1993) and b) validation (1994-2010) periods.

Figure 6. Spatial distribution of: a) simulated, and b) observed DJF precipitation (mm) during the validation period (1994-2010). c) Spatial distribution of the difference (%) between these two fields.

Figure 7. Spatial distribution of the correlation coefficients between observed DJF precipitation and predicted one by the SD model for each grid point during: a) 1950-2010 recalibration, and b) 1971-2000 periods. c) Difference in percentage the between the observed DJF precipitation and the SD modeled one for the period 1971-2000.

Figure 8. Differences (%) between the SD precipitation from 20 GCMs and the observed DJF precipitation for the 1971-2000 period. The areas where the differences are significant at the 95% confidence level (according to the Wilcoxon-Mann-Whitney non-parametric rank sum test) are marked by gray dots, and the numbers in brackets represent the percentages of these areas.

Figure 9. As in Figure 8, but for direct precipitation outputs of the 20 GCMs.

Figure 10. Changes (%) in projected (2071-2100) DJF precipitation compared to the present (1971-2000) SD precipitation for each GCM under the RCP2.6 scenario. The areas where the differences are significant at the 95% confidence level (according to the Wilcoxon-Mann-Whitney non-parametric rank sum test) are marked by gray dots, and the numbers in brackets represent the percentages of these areas.

Figure 11. As in Figure 10, but for the RCP4.5 scenario.

Figure 12. As in Figure 10, but for the RCP8.5 scenario.

Figure 13. Percentage of 20 SD GCMs that predict a positive or negative change in projected (2071-2100) DJF precipitation respect to the present (1971-2000) for each grid point, under: a) RCP2.6, b) RCP4.5, and c) RCP8.5 scenarios. The positive or negative sign of the percentage corresponds to an increase or decrease, respectively, in the projected change, with a coherence value higher than 55%.

Figure 14. Changes (%) in projected (2071-2100) DJF precipitation compared to the present (1971-2000) SD precipitation for the ensemble multi-model under the: a) RCP2.6, b) RCP4.5, and c) RCP8.5 scenarios. The areas where the differences are significant at the 95% confidence level (according to the Wilcoxon-Mann-Whitney non-parametric rank sum test) are marked by gray dots, and the numbers in brackets represent the percentages of these areas with positive (P), negative (N) and total (A) change.

**Table caption**

Table 1. CMIP5 models used for the analysis of SD at both present climate (1971-2000), and future climate (2071-2100) under the RCP2.6, RCP4.5 and RCP8.5 scenarios.

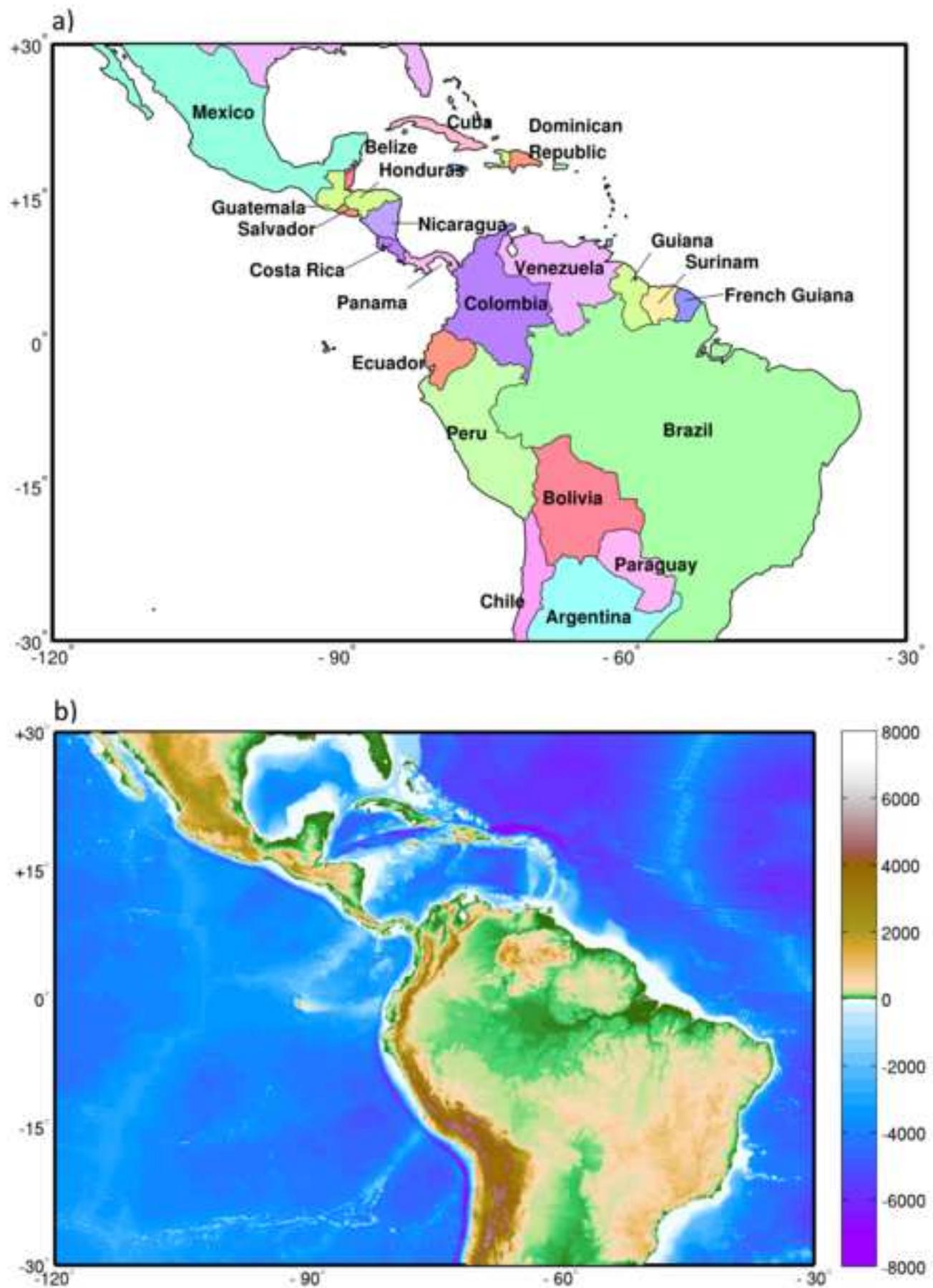

Figure 2 
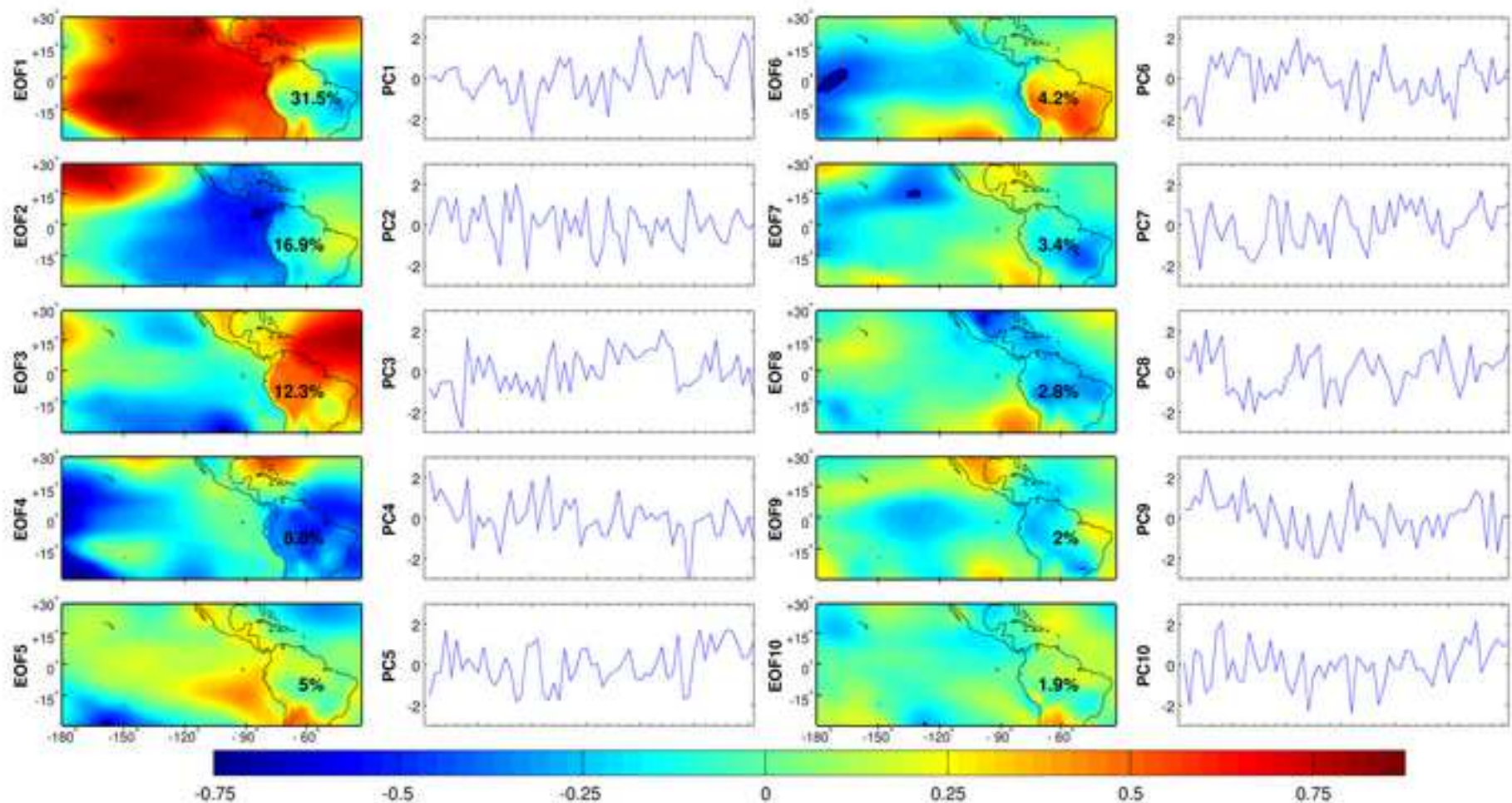

Figure 3

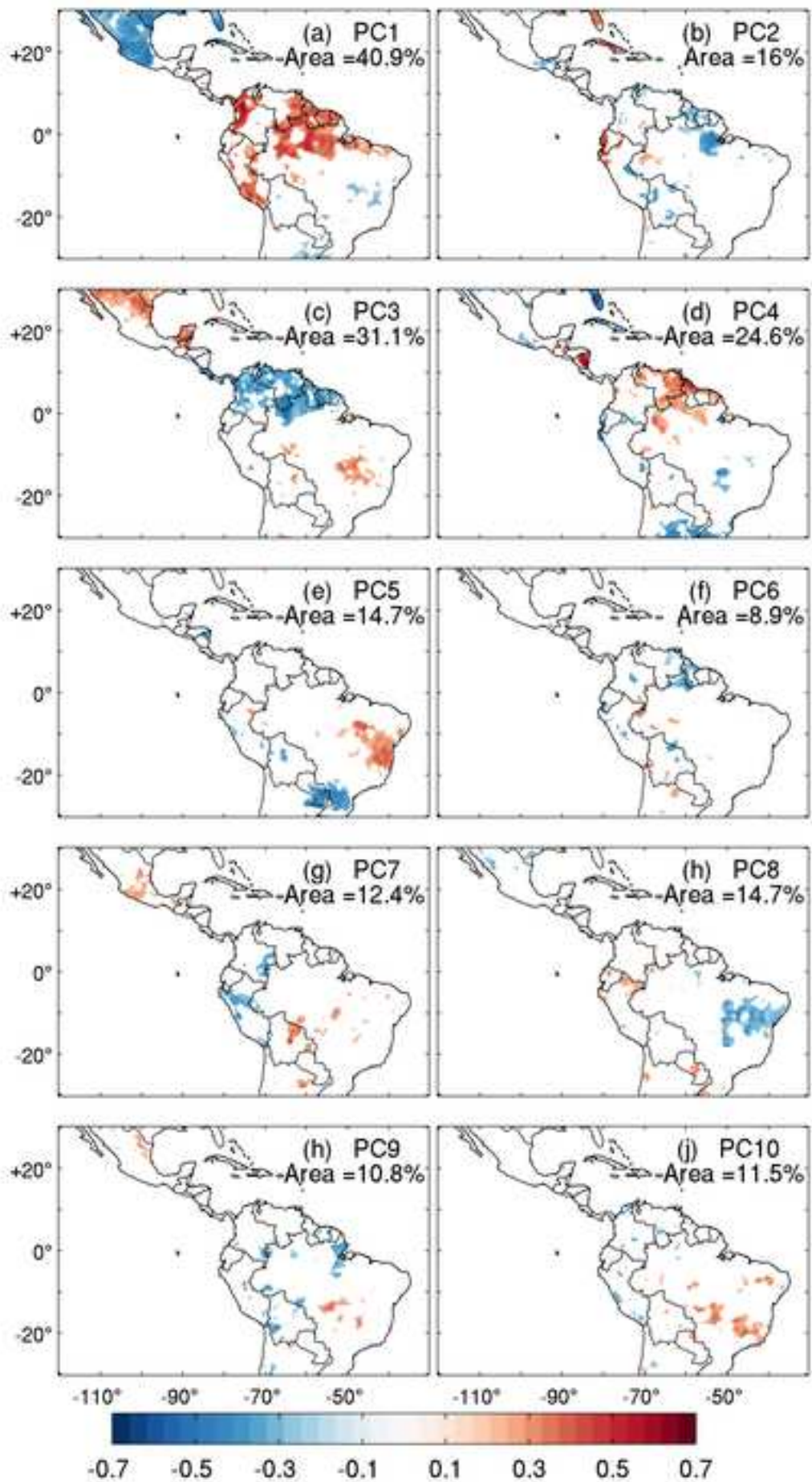

Figure 4

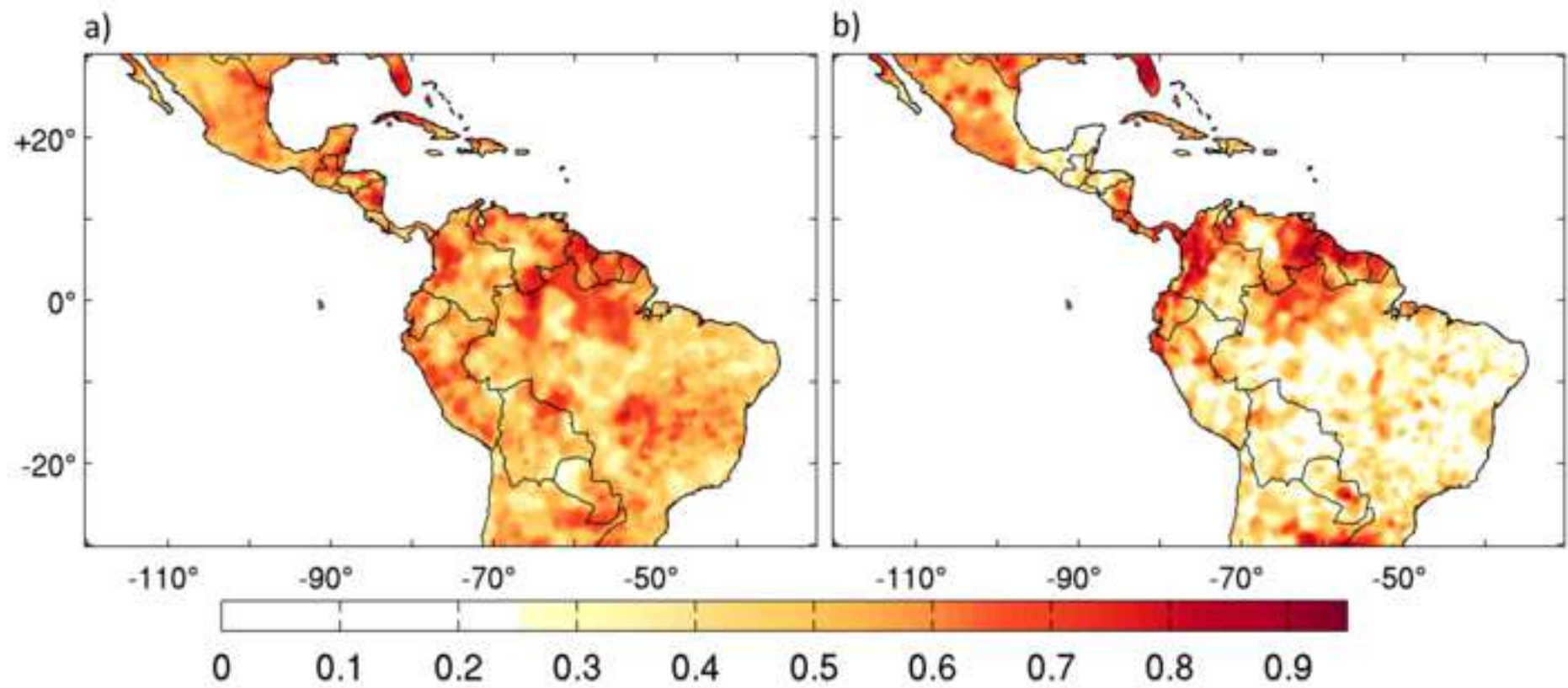

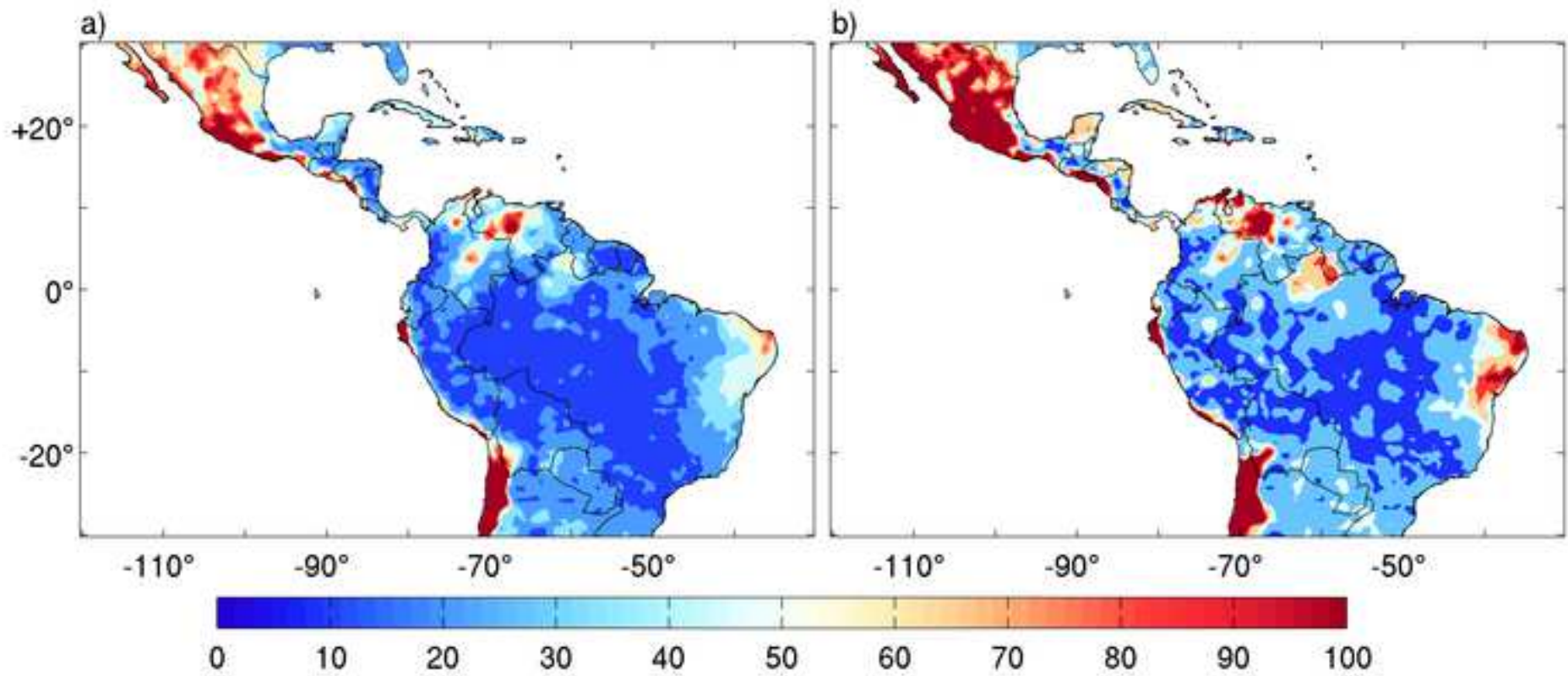

Figure 6

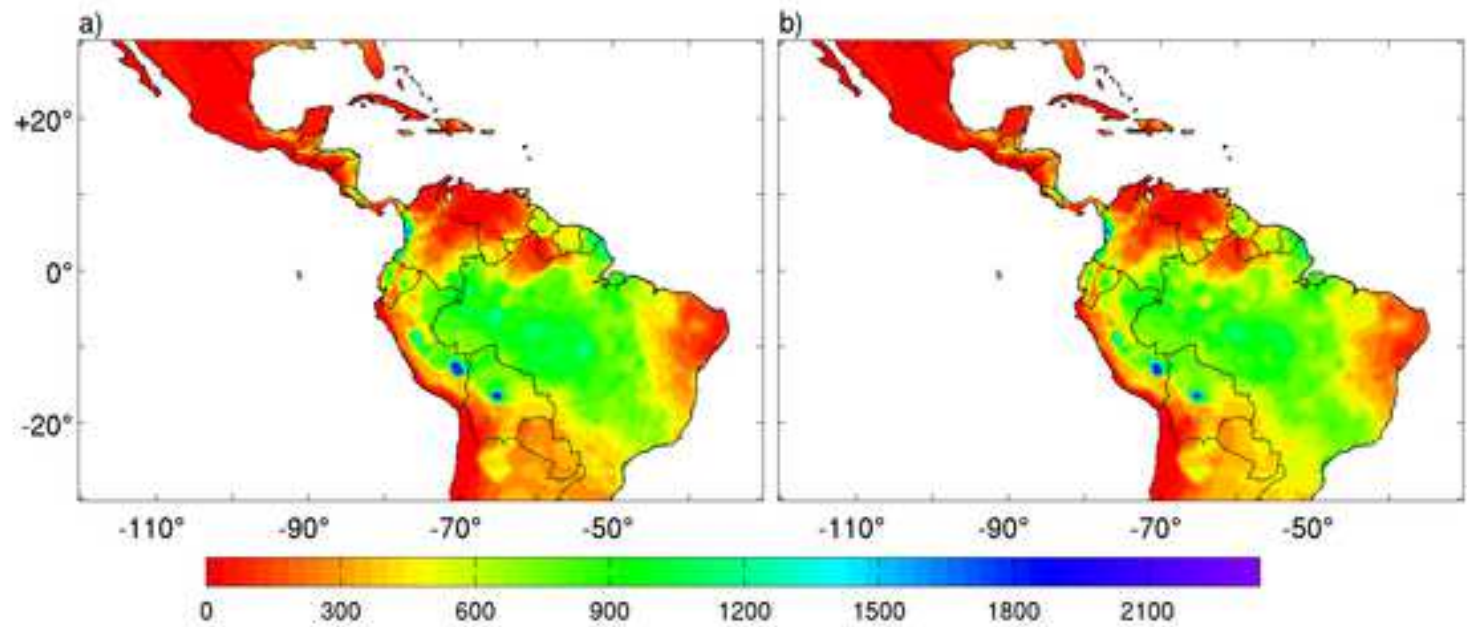
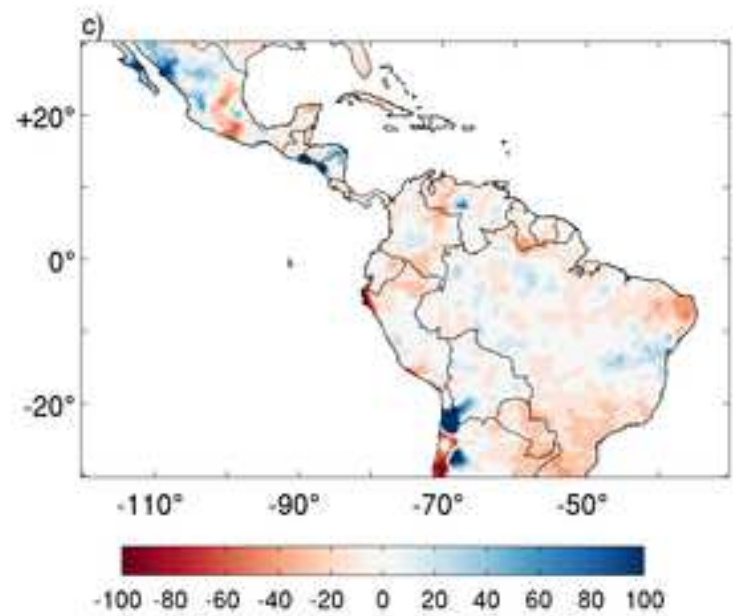

Figure 7

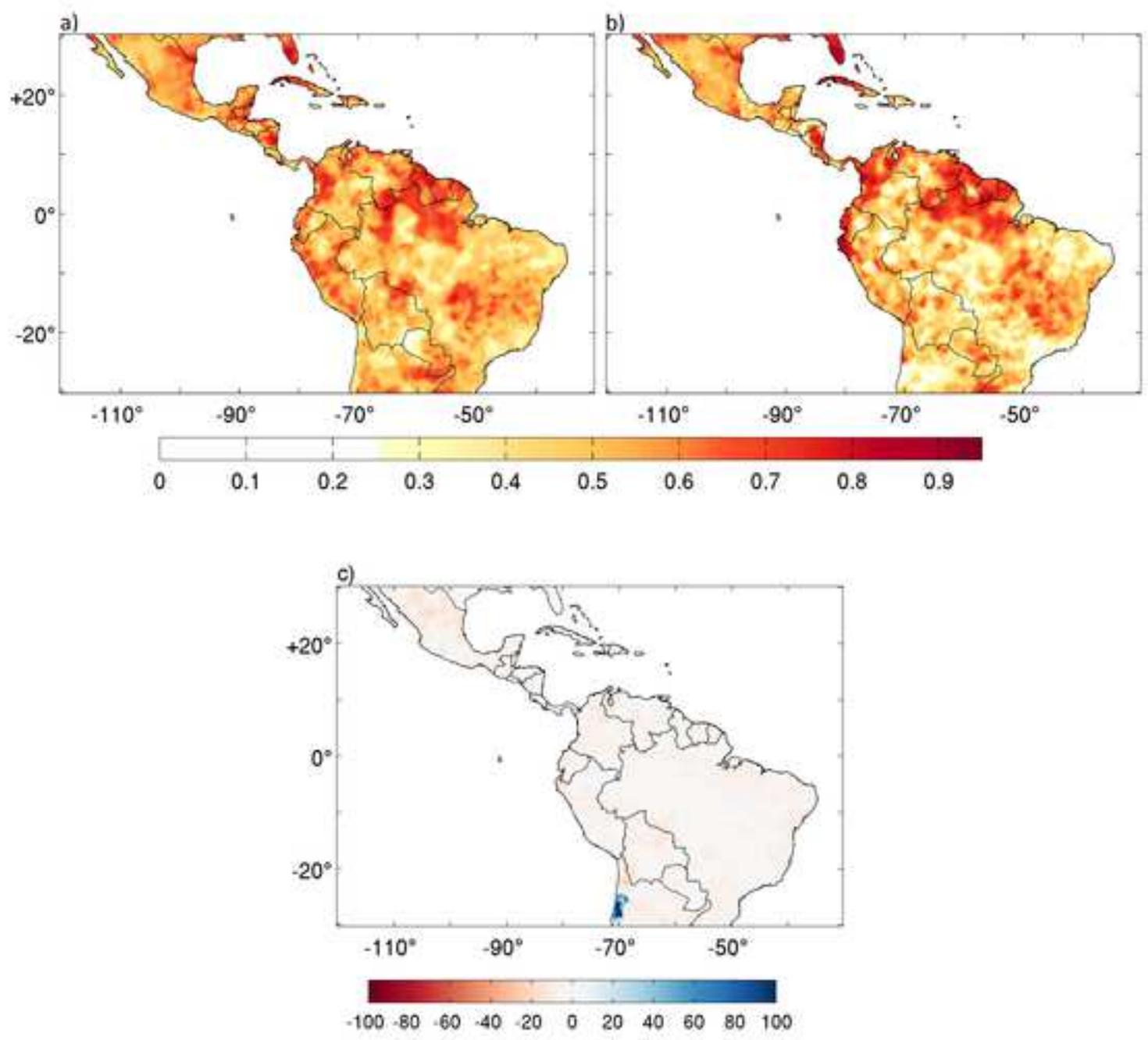

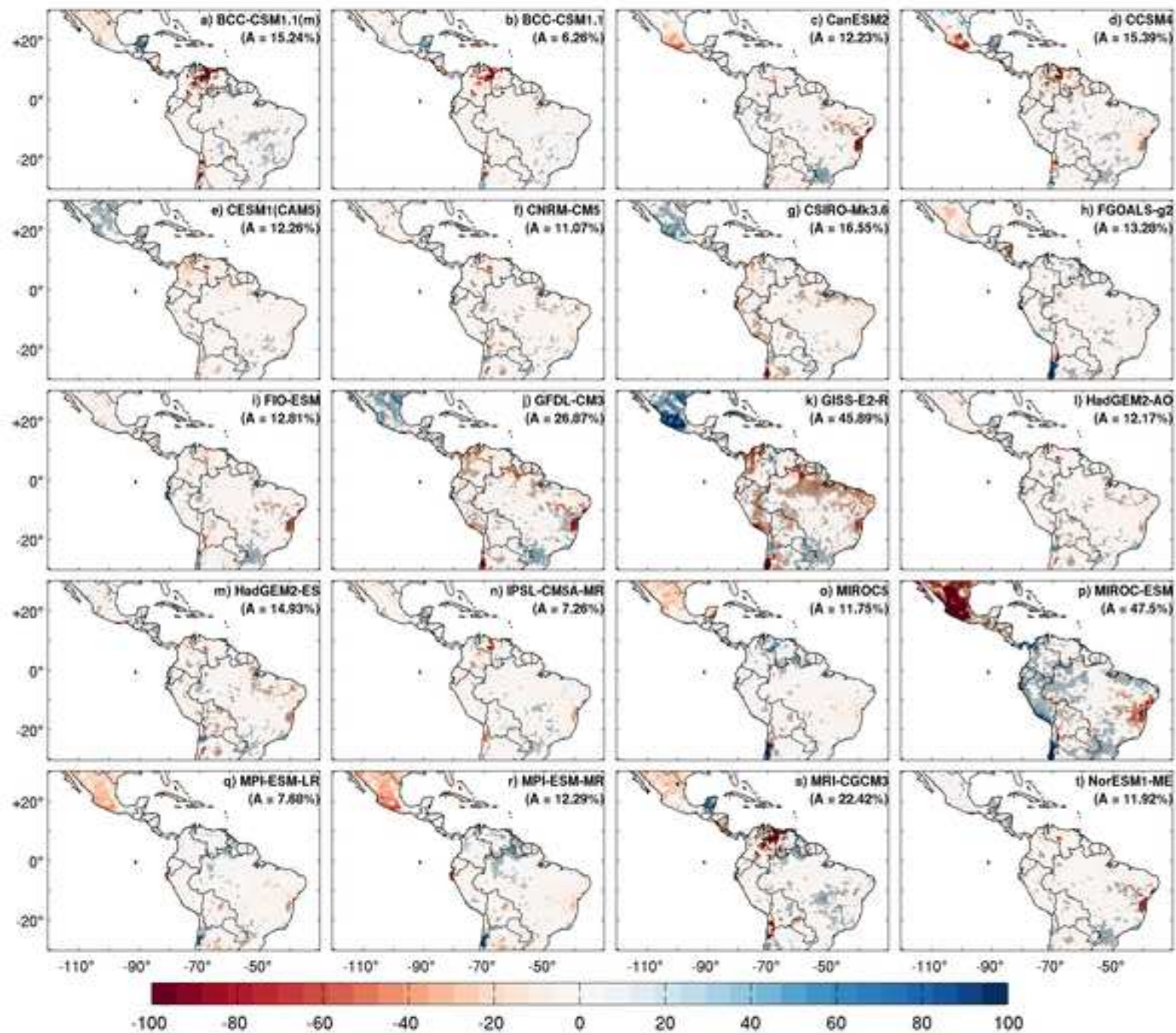

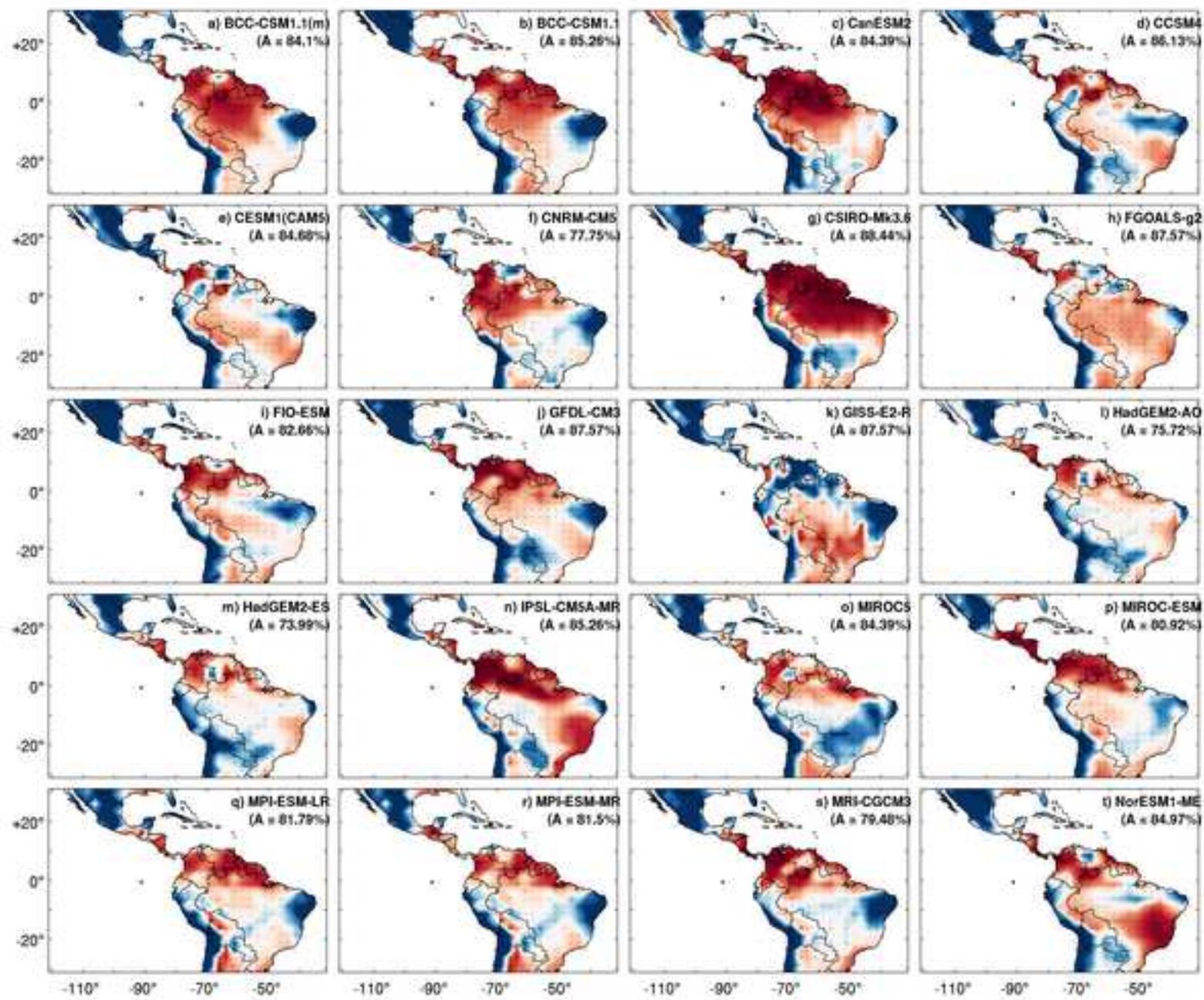

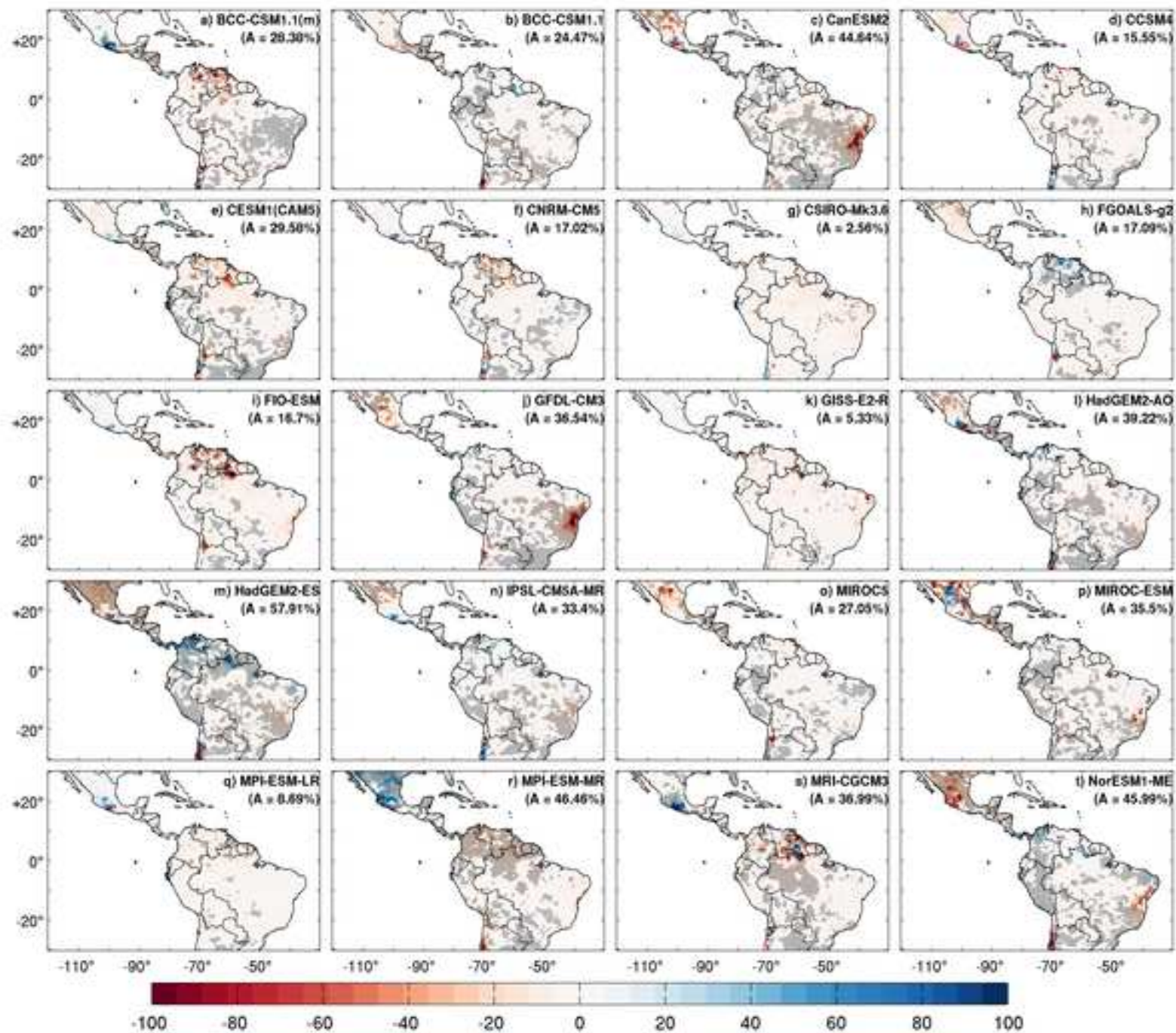

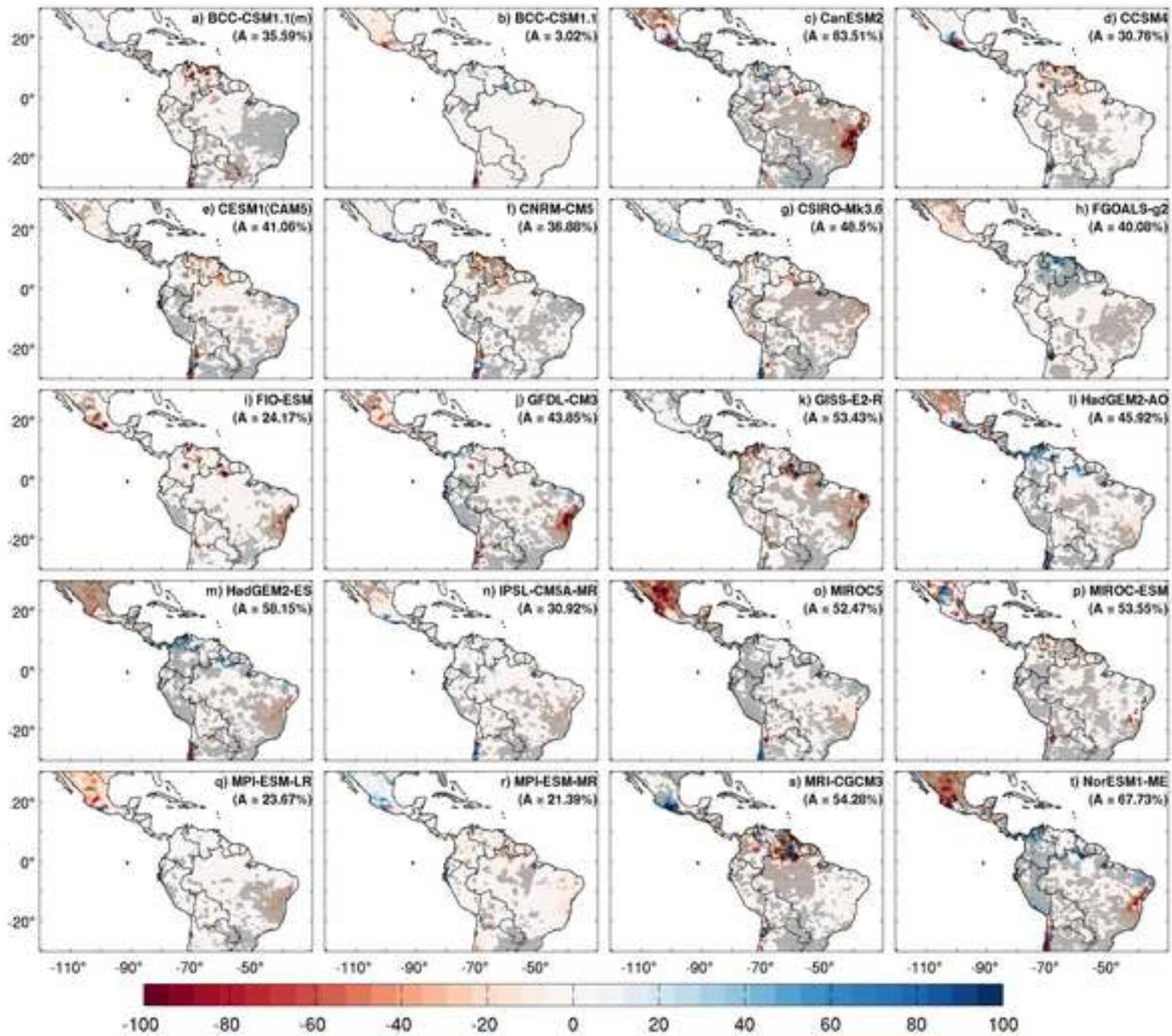

Figure 12

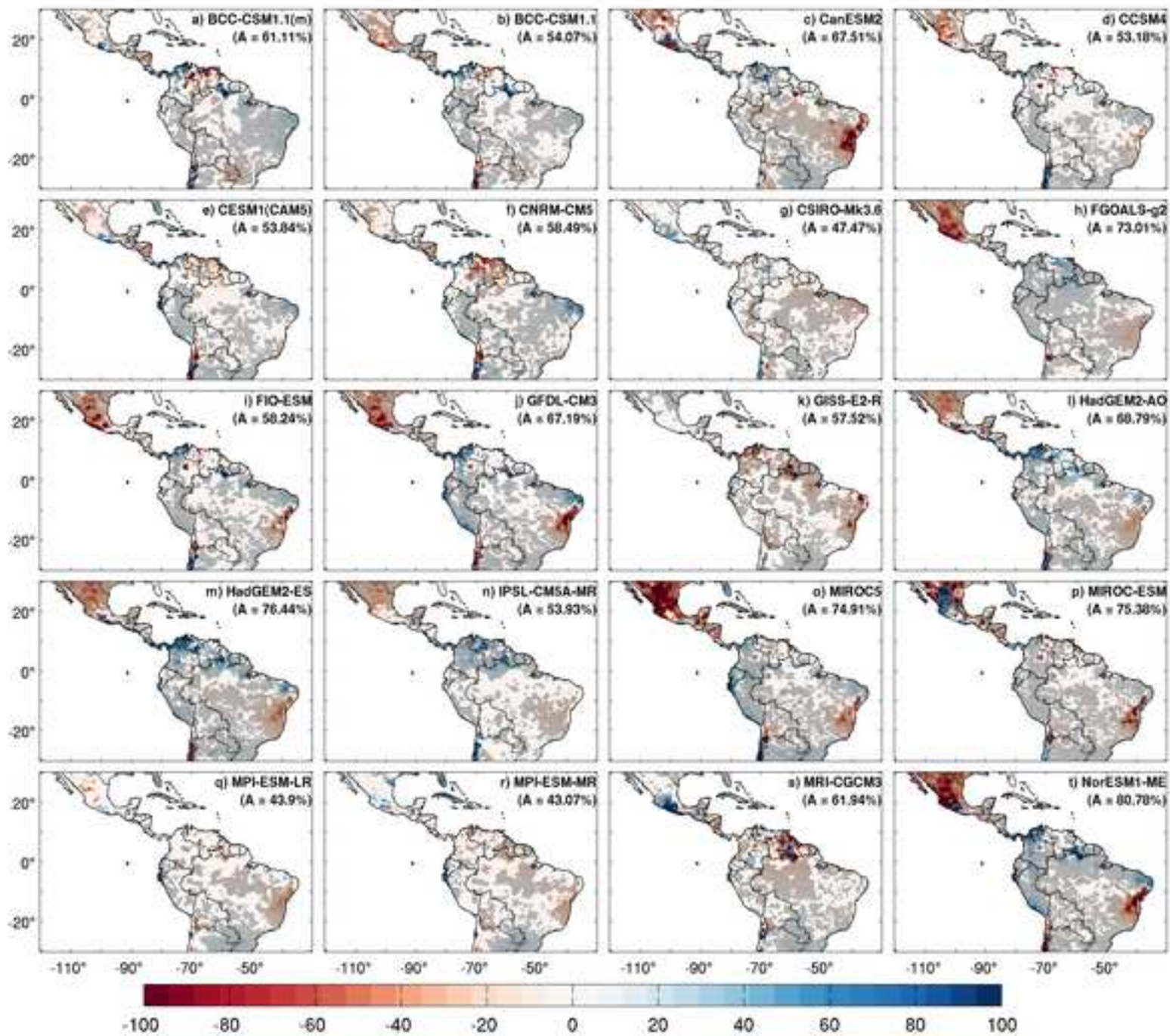

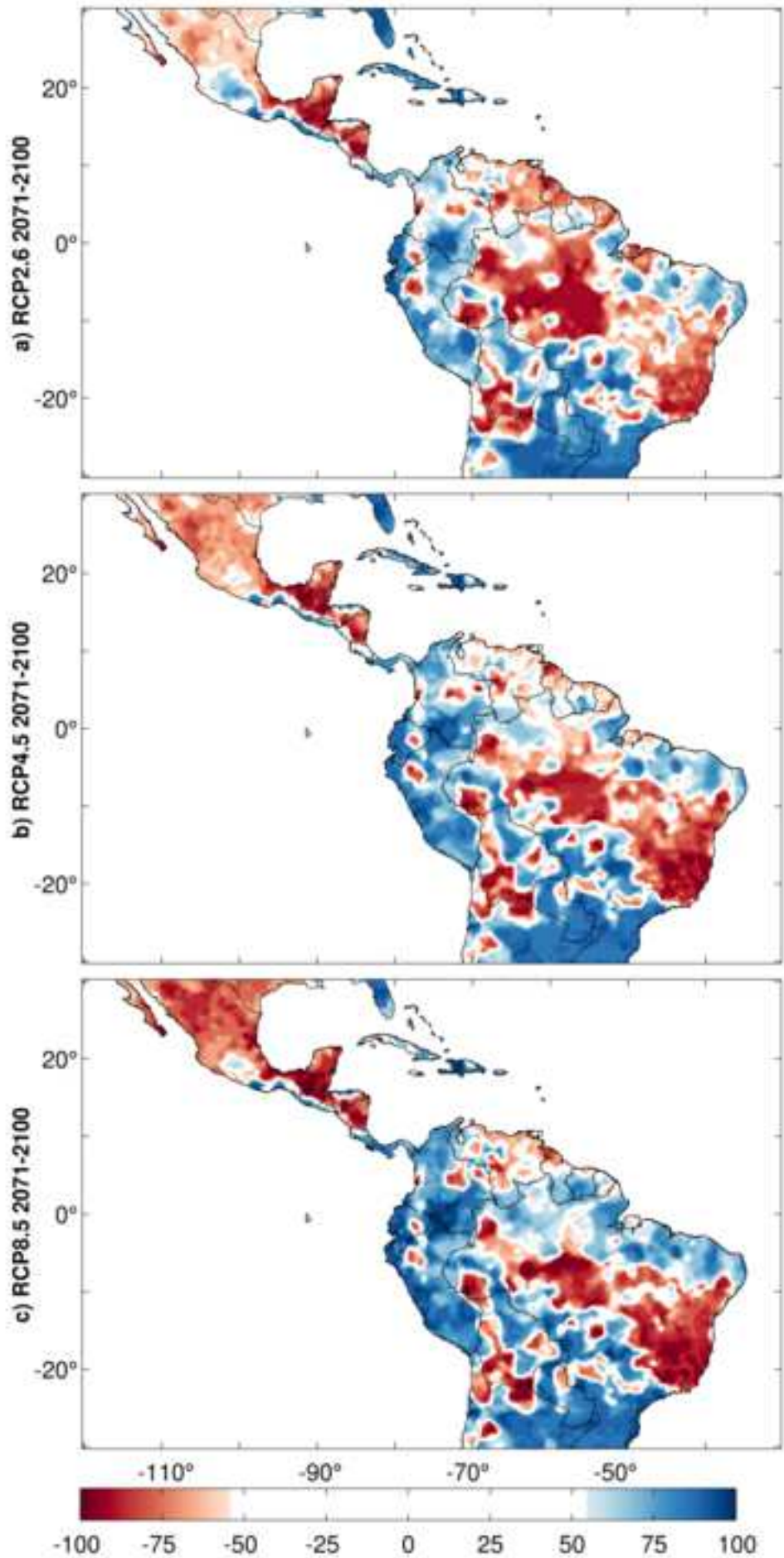

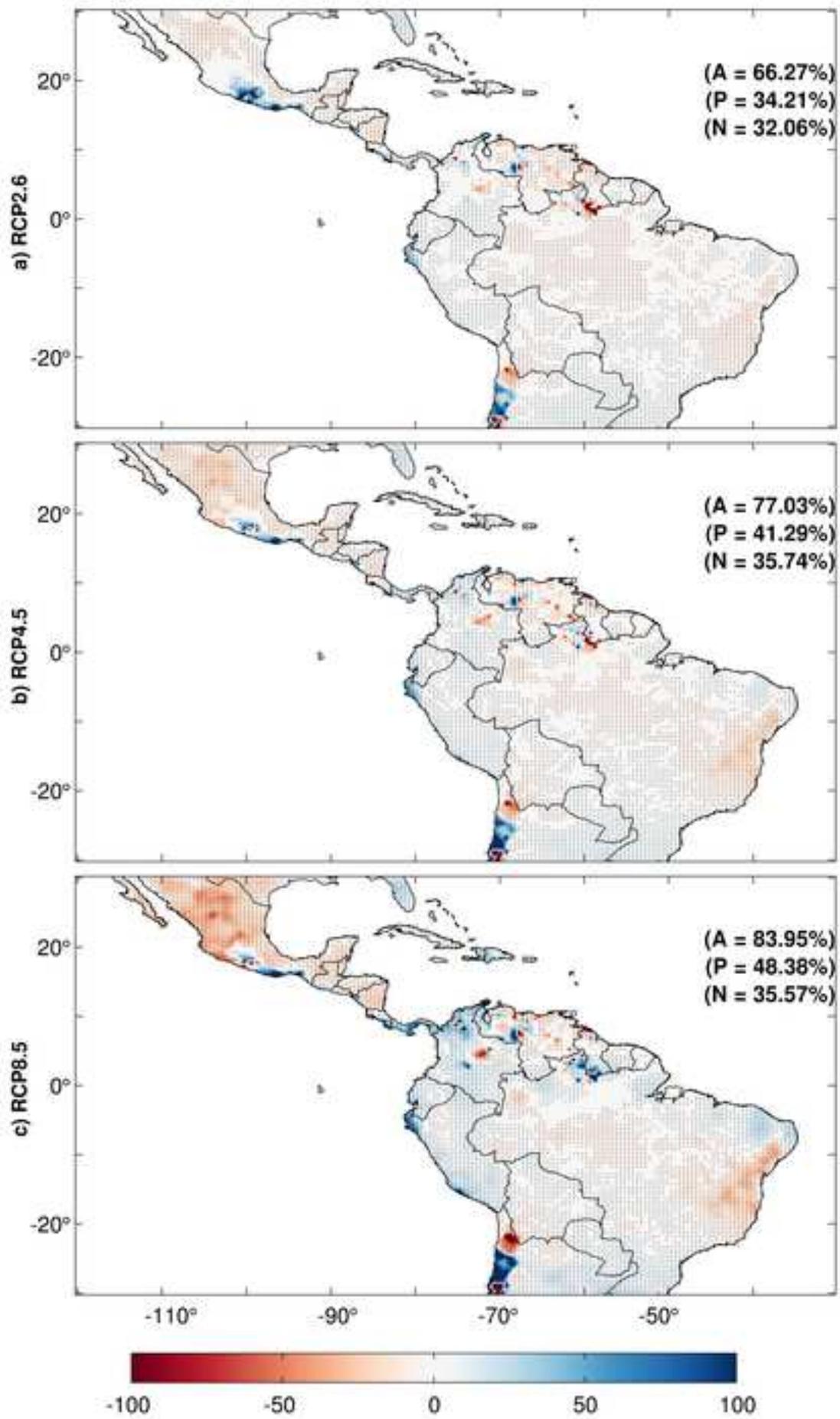

Table 1. CMIP5 models used for the analysis of SD at both present climate (1971-2000), and future climate (2071-2100) under the RCP2.6, RCP4.5 and RCP8.5 scenarios.

| Label | GCM | Centre | Label | GCM | Centre |
|---|---|---|---|---|---|
| a | BCC-CSM1.1(m) | Beijing Climate Center, China Meteorological Administration (BCC/China) | k | GISS-E2-R | NASA Goddard Institute for Space Studies (NASA GISS/USA) |
| b | BCC-CSM1.1 | | l | HadGEM2-AO | National Institute of Meteorological Research (NIMR/South Korea) |
| c | CanESM2 | Canadian Centre for Climate Modeling and Analysis (CCCma/Canada) | m | HadGEM2-ES | Met Office Hadley Centre (MOHC/UK) |
| d | CCSM4 | National Center for Atmospheric Research (NCAR/USA) | n | IPSL-CM5A-MR | Institute Pierre-Simon Laplace (IPSL/France) |
| e | CESM1(CAM5) | National Center for Atmospheric Research (NSF-DOE NCAR/USA) | o | MIROC5 | National Institute for Environmental Studies, The university of Tokyo (MIROC/Japan) |
| f | CNRM-CM5 | Centre National de Recherches Meteorologiques / Centre Europeen de Recherche et Formation Avancees en Calcul Scientifique (CNRM/France) | p | MIROC-ESM | Japan Agency for Marine-Earth Science and Technology (JAMSTEC), The University of Tokyo Atmosphere Ocean Research Institute (AORI) and National Institute for Environmental Studies (NIES) |
| g | CSIRO-Mk3.6 | Communication Scientific and Industrial Research Organization (CSIRO/Australia) | q | MPI-ESM-LR | Max Planck Institute for Meteorology (MPI-M/Germany) |
| h | FGOALS-g2 | LASG, Institute of Atmospheric Physics, Chinese Academy of Sciences; and CESS, Tsinghua University | r | MPI-ESM-MR | |
| i | FIO-ESM | The First Institute of Oceanography, SOA, China | s | MRI-CGCM3 | Meteorological Research Institute (MRI/Japan) |
| j | GFDL-CM3 | NOAA Geophysical Fluid Dynamics Laboratory (GFDL/USA) | t | NorESM1-ME | Norwegian Climate Centre (NCC/Norway) |